\documentclass{PoS}

\title{The connection between the Galactic halo and ancient Dwarf Satellites} 
\ShortTitle{The connection between the Galactic halo and ancient Dwarf Satellites} 
\author{\speaker{Emanuele Spitoni}\\
        Dipartimento di Fisica,  Universit\`a di Trieste\\
        E-mail: \email{spitoni@oats.inaf.it}}
\author{Fiorenzo Vincenzo\\
        Dipartimento di Fisica,  Universit\`a di Trieste\\
        E-mail: \email{vincenzo@oats.inaf.it}}
\author{Francesca Matteucci\\
         Dipartimento di Fisica,  Universit\`a di Trieste\\
        E-mail: \email{matteucci@oats.inaf.it}}
\author{Donatella Romano\\
         I.N.A.F. Osservatorio Astronomico di Bologna\\
        E-mail: \email{donatella.romano@oabo.inaf.it}}

\abstract{
 We explore the hypothesis that the classical and
ultra-faint dwarf spheroidal satellites of the Milky Way have been the
building blocks of the Galactic halo by comparing their [O/Fe]
and [Ba/Fe] versus [Fe/H] patterns with the ones observed in Galactic
halo stars. Oxygen abundances deviate substantially from the
observed abundances in the Galactic halo stars for [Fe/H] values
larger than -2 dex, while they overlap for lower metallicities. On the
other hand, for the [Ba/Fe] ratio the discrepancy is extended at all
[Fe/H] values, suggesting that the majority of stars in the halo are
likely to have been formed in situ.  Therefore, we suggest that
[Ba/Fe] ratios are a better diagnostic than [O/Fe] ratios.
Moreover, we show the effects of an enriched
infall of gas with the same chemical abundances as the matter ejected
and/or stripped from dwarf satellites of the Milky Way on the chemical
evolution of the Galactic halo. We find that the resulting chemical
abundances of the halo stars depend on the assumed infall time scale,
and the presence of a threshold in the gas for star formation. }

\FullConference{Frontier Research in Astrophysics - II\\
         23-28 May 2016\\
         Mondello (Palermo), Italy}

\begin{document}

\section{Introduction}

 According to the $\Lambda$CDM paradigm, a Milky Way-like galaxy must have
formed by the merging of a large number of smaller systems. In
particular, dwarf spheroidal galaxies (dSphs) were proposed in the
past as the best candidate small progenitor systems, which merged
through cosmic time to eventually form the stellar halo component of
the Galaxy (e.g. Grebel 2005).

 On the other hand, Fiorentino et al. (2015) using RR Lyrae stars as
tracers of the Galactic halo ancient stellar component, showed that
dSphs do not appear to be the major building-blocks of the
halo. Leading physical arguments suggest an extreme upper limit of
50\% to their contribution.

 In recent years, the Sloan Digital Sky Survey (SDSS, York et
 al. 2000) were able to discover an entirely new population of
 hitherto unknown stellar systems: the so-called ultra faint dwarf
 spheroidal galaxies (UfDs), which are characterized by extremely low
 luminosities, high dark matter content, and very old and iron-poor
 stellar populations (Belokurov et al. 2006; Norris et al (2008,
 2010); Brown et al 2012).   The number of UfDs has
 increased constantly in the last decade and completeness estimates
 suggest that many more of these faint satellites are still to be
 discovered in the Local Group (Tollerud et al. 2008). This fact might
 place them as the survived building blocks of the Galaxy stellar
 halo, dramatically lacking in the past.

In Spitoni et al. (2016) we test the hypothesis that dSph and UfD
galaxies have been the building blocks of the Galactic halo, by
assuming that the halo formed by accretion of stars belonging to these
galaxies. Moreover,  extending  the results of Spitoni (2015) to detailed
chemical evolution models in which the IRA is relaxed, we explored the scenario, in which the 
Galactic halo formed by accretion of chemically enriched gas
originating from dSph and UfD galaxies.

\section {The chemical evolution models}

\subsection{The Milky Way}
We  consider the following two reference chemical evolution models for the 
MW galaxy:

\begin{enumerate}

\item The  classical two-infall  model (2IM) presented by Brusadin  et al. (2013). The Galaxy is assumed to have
formed by means of two main infall episodes: the first formed the halo
and the thick disc, the second the thin disc. 

\item The two-infall model plus outflow of Brusadin et al. (2013; here we  indicate it as the 2IMW model). In this model  a gas outflow occuring
during the halo phase with a rate proportional to the star formation
rate through a free parameter is considered.  
  Following Hartwick (1976), the outflow rate is defined as:
\begin{equation}
\frac{d \sigma_w}{dt}=-\omega \psi(t),
\end{equation}
where $\omega$ is the outflow efficiency.

\end{enumerate}

In Table 1 the principal characteristics of the two chemical evolution
models for the MW are summarized: in the second  column the
time-scale $\tau_H$ of halo formation, in the third  the time-scale $\tau_D$ of
the thin disc formation, are drawn. The adopted threshold in the
surface gas density for the star formation (SF) is reported in columns
4. In column 5 the exponent of the Schmidt (1959) law is shown, in
columns 6 and 7 we report the star formation efficiency and the IMF,
respectively. Finally, in the last column the presence of the wind is
indicated in term of the efficiency $\omega$.

\begin{figure}[h]
  \includegraphics[scale=0.35]{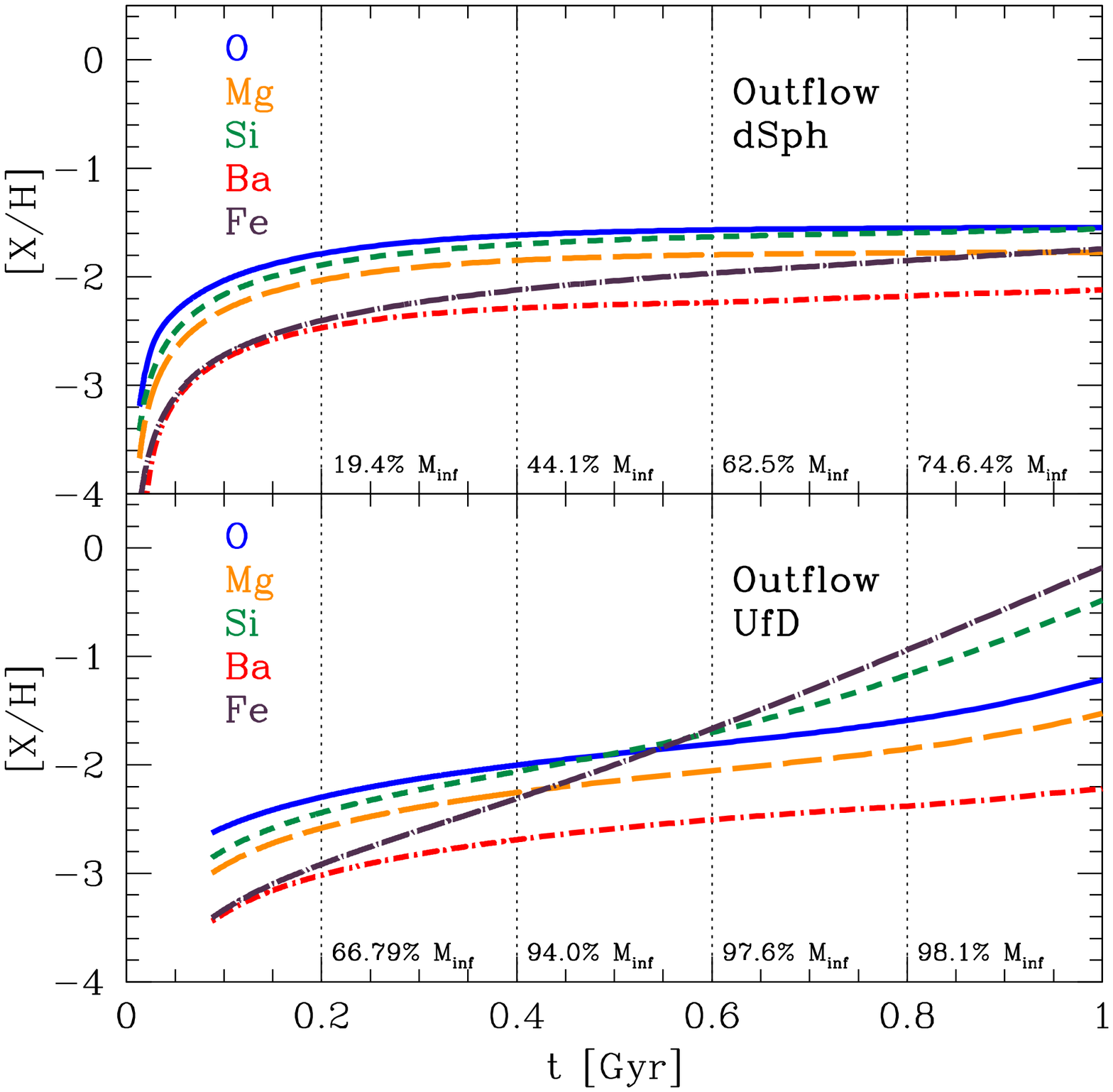}
  \includegraphics[scale=0.35]{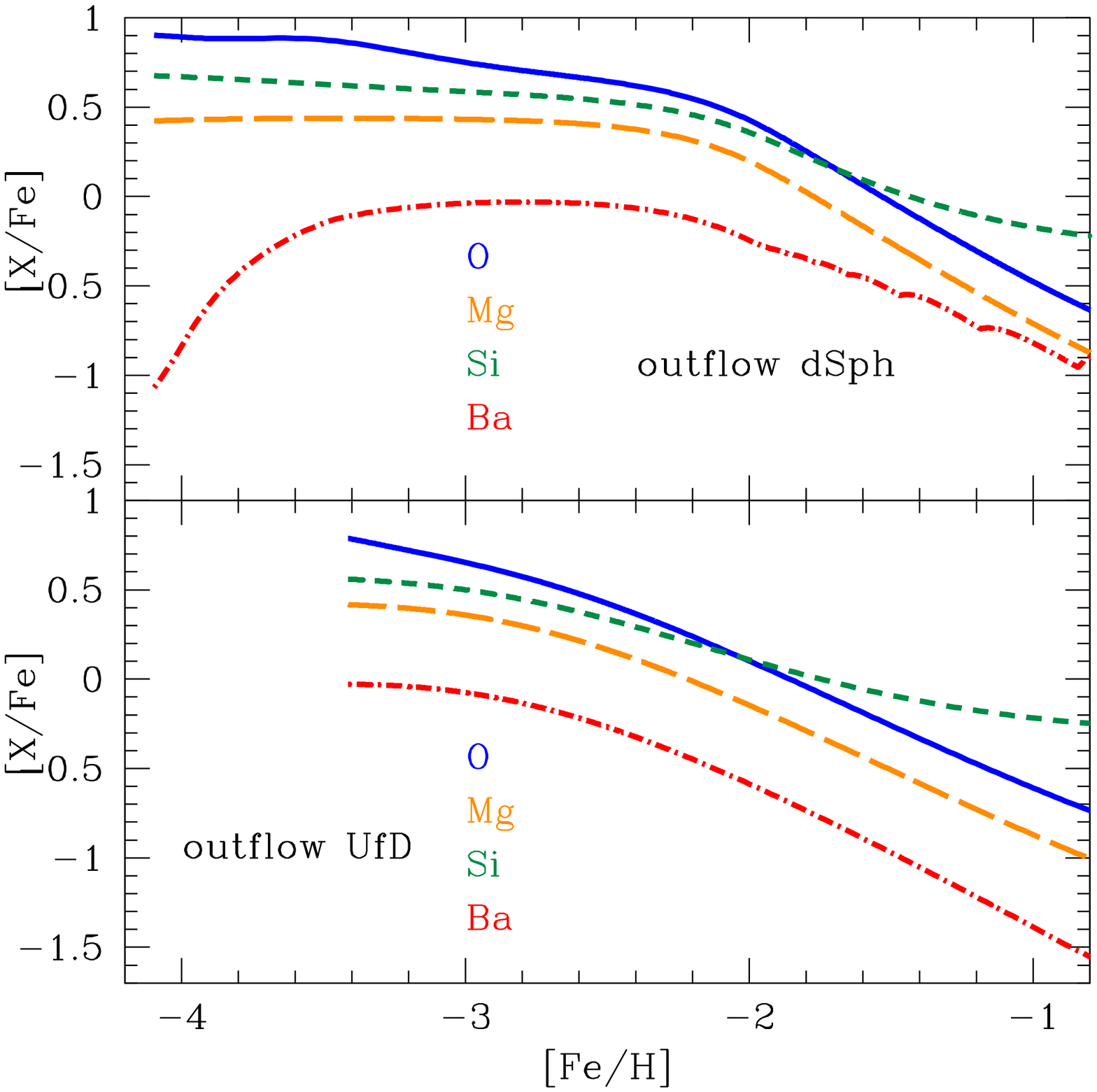}
  \caption{{\it Left panels}:  The evolution in time of the chemical abundances for  O, Mg, Si,  Ba, Fe in  the gas ejected  as galactic wind from dSphs and UfDs. The onset of the wind happens at later times compared with  dSph objects.  We also indicate the cumulative ejected gas mass by outflows at 0.2, 0.4, 0.6, and 0.8 Gyr in terms of percentage of the infall mass $M_{inf}$. 
{\it Right panels}: The abundance ratio [X/Fe] as a function of [Fe/H] for the following chemical elements: O, Mg, Si, and Ba  of the outflowing gas ejected by a  dSph galaxy, and by a UfD galaxy.}
\label{wind1}
\end{figure}

\begin{table*}

%\normalsize 
\label{TMW}
\begin{center}
\begin{tabular}{c c c c c c c c c c c c }
  \hline
\multicolumn{9}{c}{{\it \normalsize The Milky Way: the solar neighborhood  model parameters}}\\
\\
%\noalign{\smallskip}

 Models &$\tau_H$& $\tau_D$&Threshold & $k$& $\nu$& IMF& $\omega$&\\

&  [Gyr]& [Gyr]& [M$_{\odot}$pc$^{-2}$]&&[Gyr $^{-1}$] &&[Gyr$^{-1}$]&\\  
  
%\noalign{\smallskip}
\hline
%\noalign{\smallskip}

2IM & 0.8   & 7 & 4 (halo-thick disc) &1.5 & 2 (halo-thick disc)& Scalo (1986)&/\\
&&&7 (thin disc)&& 1 (thin disc)\\
%\noalign{\smallskip}
%\hline
%\noalign{\smallskip}
\hline

2IMW &  0.2   & 7 & 4 (halo-thick disc) &1.5 & 2 (halo-thick disc)& Scalo (1986)&14\\
&&&7 (thin disc)&& 1 (thin disc)\\

\hline
%3IM&  3 infall & 0.2   & 1.25&6  &4 (halo)&1.5&2  & Scalo (1986)&/\\
%&&&&& 5 (thick disc)&&10\\
%&&&&& 7 (thin disc)&&1 \\
%\noalign{\smallskip}
% \hline
%\noalign{\smallskip}
\end{tabular}
\end{center}
\caption{Parameters of the chemical evolution models for the Milky Way (Spitoni al. 2016) in the solar neighborhood.}
\end{table*}

\begin{table*}
\begin{tabular}{c c c c c c c c c c c}
\hline

Models&$\nu$& $k$& $\omega$ & $\tau_{inf}$  & $M_{inf}$ & $M_{DM}$ & $r_{L}$ & $S=\frac{r_{L}}{r_{DM}}$ & IMF&\\

&$[\mbox{Gyr}^{-1}$] && & [Gyr]  & [$M_{\odot}$] & [$M_{\odot}$] & [pc] & & & \\

\hline

dSphs&0.1 & 1&10 & $0.5$  & $10^{7}$ & $3.4\cdot10^{8}$ & $260$ & $0.52$ & Salpeter(1955)&  \\
\\
Ufds&0.01 & 1&10 & $0.001$  & $10^{5}$ & $10^{6}$ & $35$ & $0.1$ & Salpeter (1955)&\\
\hline
\end{tabular}
\caption{ Parameters of the chemical evolution model for general dSph and UfD  galaxies (Spitoni et al. 2016). }
\end{table*}

\subsection{The dSph and UfD  galaxies}

  In Table 2  the main
parameters of generic models for ``classical'' dSph and UfD galaxies
are reported, respectively. The star formation efficiency $\nu$, the
exponent $k$ of the Kennicutt (1998) law, and the wind efficiency
$\omega$ are drawn in column one, two and three, respectively. In the
other columns are reported: the infall timescale (column 4);  total infall gas mass (column 5); mass
of the dark matter halo (column 6); effective radius of the luminous
(baryonic) matter (column 7); ratio between the core radius of the DM
halo and the effective radius of the luminous matter (column 8); 
in column 9 the adopted IMF is indicated.   We assume that UfD objects are characterized by a very
small star formation efficiency (0.01 Gyr$^{-1}$) and by an extremely
short time scale of formation (0.001 Gyr). 
  We point out that in the modeling the dSphs and UfDs we did not
consider any threshold in the gas density for star formation, as in Vincenzo et al. (2014). 

The time
at which the galactic wind starts in  dSphs is at 0.013 Gyr after the
galactic formation, whereas for UfDs at 0.088 Gyr.
As expected, the  UfD galaxies develop a  wind at later times  because of the 
smaller adopted star formation efficiency (SFE).

\subsection{Nucleosynthesis prescriptions}

 We adopt the nucleosynthesis prescriptions of Romano et al. (2010, model 15), who provide a compilation of stellar yields able to reproduce several chemical abundance patterns in the solar neighborhood. 
In particular, they assume the following sets of stellar yields.

For  barium, we assume the stellar yields of Cescutti et al. (2006, model 1, table 4). 
In particular, Cescutti et al. (2006) includes the metallicity-dependent stellar yields of Ba as computed by Busso et al. (2001), in which barium is produced by low-mass AGB stars, 
with mass in the range $1.0\le M \le3.0$ M$_{\odot}$, as an s-process neutron capture element. A second channel for the Ba-production was 
included by Cescutti et al. (2006), by assuming that massive stars in their final explosive stage are capable of synthesizing Ba  as a primary 
r-process element. Such r-process Ba producers have mass in the range $12\le M \le30$ M$_{\odot}$.  

We remark on the fact that the contribution to barium from massive
stars was empirically computed by Cescutti et al. (2006), by matching
the [Ba/Fe] versus [Fe/H] abundance pattern as observed in the Galactic
halo stars.  They assumed for massive stars the iron stellar yields of
Woosley \& Weaver (1995), as corrected by Fran\c cois et al. (2004).

\section{The enriched infall of gas}

The novelty of Spitoni et al. (2016) is to take into account in a
self-consistent way time dependent abundances.  The gas infall law is
the same as in the 2IM or 2IMW models and it is only considered a time
dependent chemical composition of the infall gas mass.

We take into account the enriched infall from dSph and UfD galaxies predicted by the following 2 models:

\begin{itemize}

\item Model i): The infall of gas which forms the Galactic halo is considered primordial up to the time at which the galactic wind  in dSphs (or UfDs) starts.  After this moment, the infalling gas presents the chemical abundances of the  wind.   In Figs. 2 and 3  we refer to this model with the label ``Name of the reference model+dSph''or ``Name of the reference model+UfD''.

\item Model ii): we explore the case of a diluted infall of gas during the MW halo phase. In particular, after the galactic wind develops in the dSph (or UfD) galaxy,  
the infalling gas has a chemical composition which, by $50$ per cent,
is contributed by the dSph (or UfD) outflows; the remaining $50$ per
cent is contributed by primordial gas of a different extra-galactic
origin.   In all the successive
figures and in the text, we refer to these models with the labels
``Name of the MW model+dSph (or UfD) MIX''.  

\end{itemize}

 In the two left panels of Fig. \ref{wind1}, we show the
evolution in time of the chemical composition of the outflowing gas
from the dSph and the UfD galaxy for O, Mg, Si, Ba and Fe. It is worth noting that in the outflows from UfD galaxies the Fe and Si abundances are larger than in the outflows from dSphs. We recall that Fe is mostly produced by Type Ia SNe and Si is also produced in a non negligible amount by the same SNe. Because in our models the ratio between the time scale of formation between UfD and dSph is extremy low ($\tau_{inf}$(UfD)/$\tau_{inf}$(dSph)=$2 \times 10^{-3}$, at later times the pollution from Type IA SN  is more evident in the UfD outflow.  In the two right panels  the [$X$/Fe] versus [Fe/H] abundance patterns are presented,
where $X$ corresponds to O, Mg, Si, and Ba.

\section{The Results}

In this section, we present the results of our chemical evolution
models for the Galactic halo.

\subsection{The Results: the Galactic halo in the model 2IM}

\begin{figure}[h]
\centering	  \includegraphics[scale=0.35]{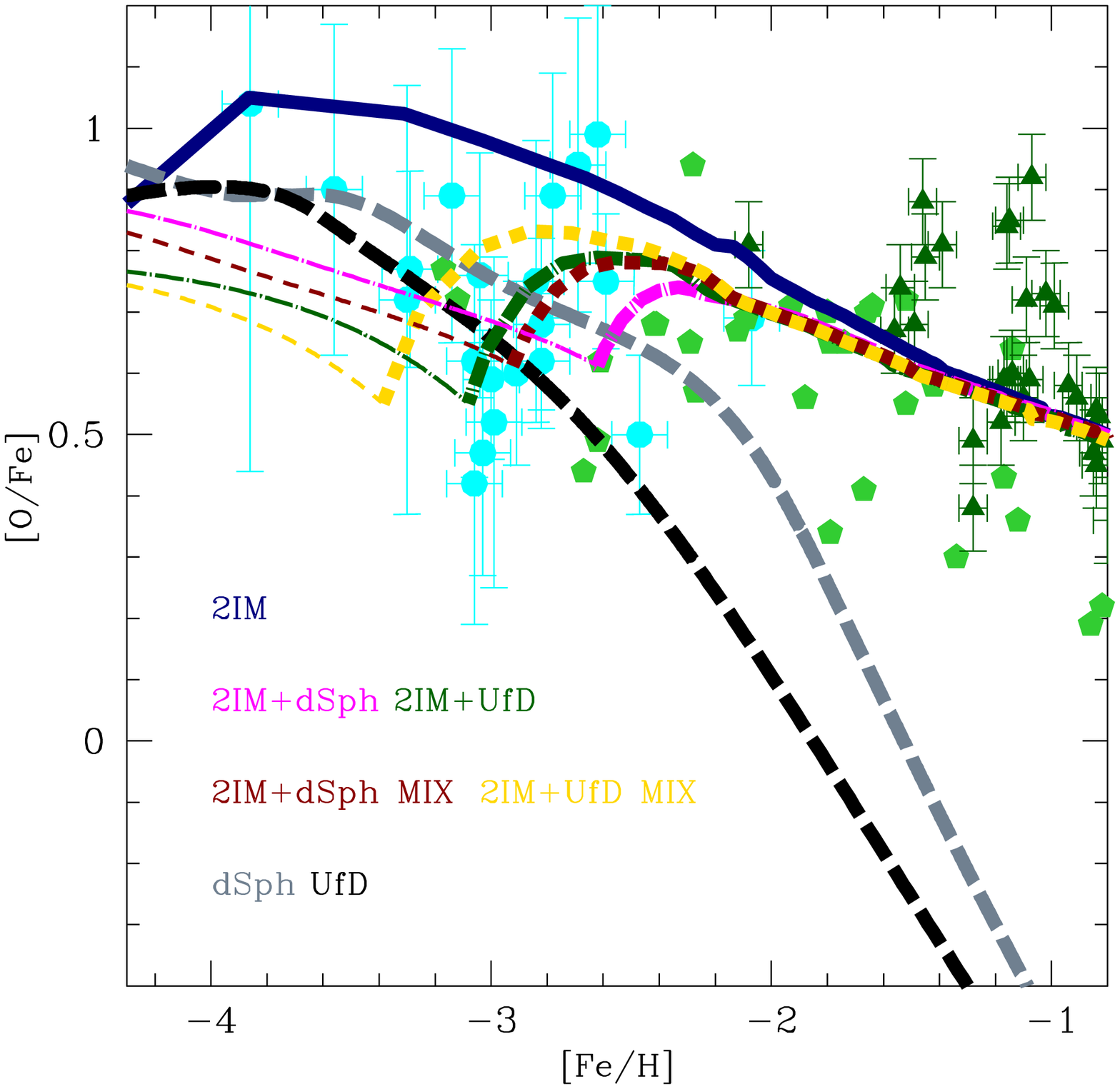}
  \centering \includegraphics[scale=0.35]{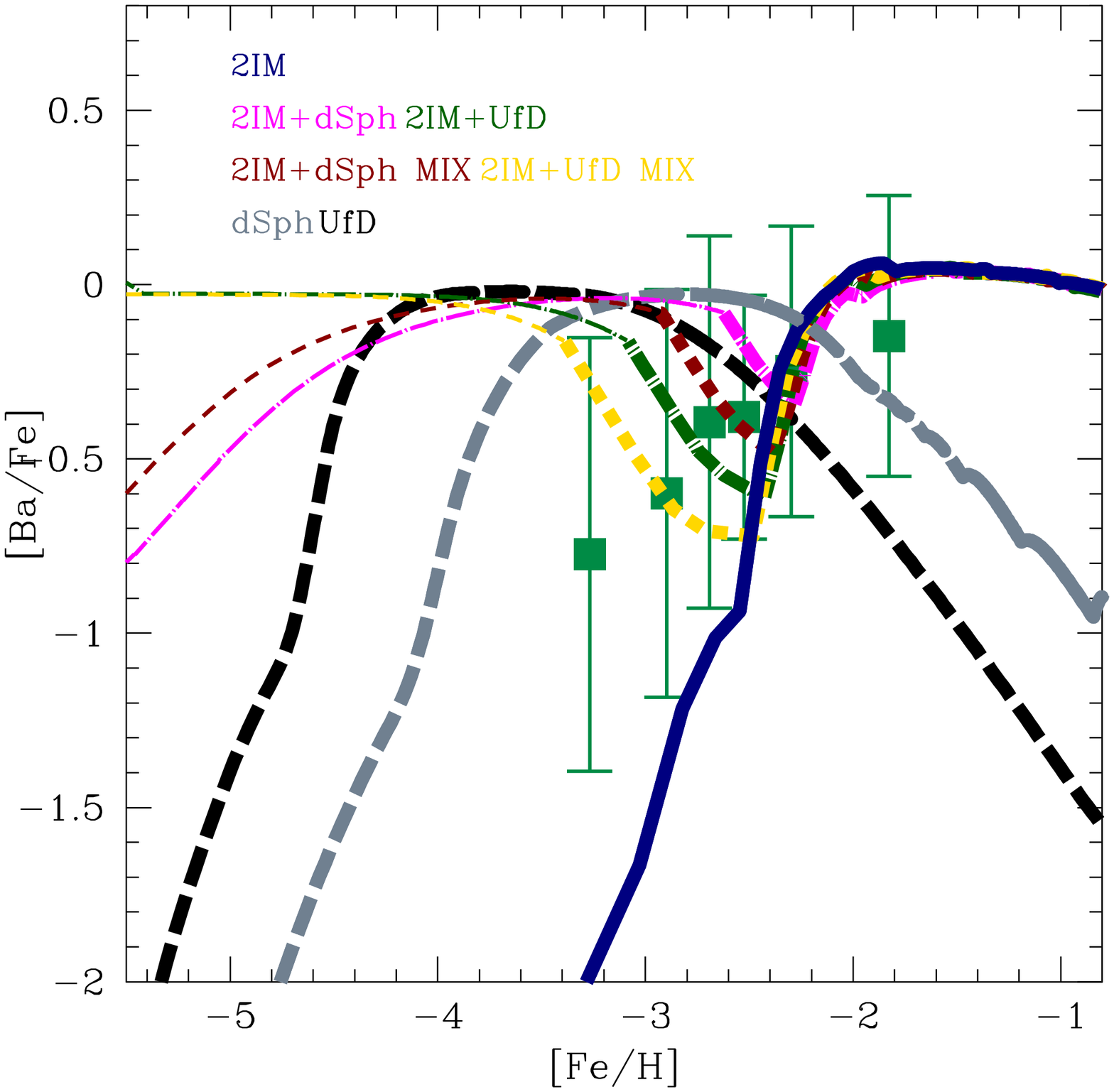}
\caption{The abundance ratios [O/Fe] vs [Fe/H] (left panel) and [Be/Fe] vs [Fe/H]  (right panel)  in the solar
neighborhood for the reference model 2IM are drawn with the solid blue
line. {\it Models of the Galactic Halo with the enriched infall from
dSph}: the magenta dashed dotted line and the red short dashed line
represent the models 2IM+dSph and 2IM+dSph MIX, respectively. {\it
Models of the Galactic Halo with the enriched infall from UfDs}: the
green dashed dotted line and the yellow short dashed line represent
the models 2IM+UfD and 2IM+UfD MIX, respectively.   Thinner lines
indicate the ISM chemical evolution phases in which the SFR did not
start yet in the Galactic halo, and during which stars are no
created.  {\it Models of the dSph and UfD galaxies}: The long dashed
gray line represents the abundance ratios for the dSph galaxies,
whereas long dashed black line for the UfD galaxies. {\it
Observational Oxygen data  of the Galactic Halo:} Cayrel et al. (2004) (cyan
circles), Akerman et al.  (2004) (light green pentagons), Gratton et
al. (2003) (dark green triangles). 
 {\it Observational  Barium data of the Galactic Halo:} Frebel  (2010).
 }
\label{O1}
\end{figure}

In the left panel of Fig. \ref{O1}, the predicted [O/Fe] versus [Fe/H] abundance
patterns are compared with the observed data in Galactic halo
stars.  In order to
directly test the hypothesis that Galactic halo stars have been
stripped from dSph or UfD systems, we show the predictions of chemical
evolution models for a typical dSph and UfD galaxy (long dashed lines
in grey and black, respectively).  The two models cannot explain the
[$\alpha$/Fe] plateau which Galactic halo stars exhibit for
$\mathrm{[Fe/H]}\ge-2.0$ dex; in fact, halo stars have always larger
[O/Fe] ratios than dSph and UfD stars.

Moreover, in the left panel of Fig. \ref{O1} we show the effects of the enriched infall
with chemical abundances taken by the outflowing gas from dSph and Ufd
objects on the [O/Fe] versus [Fe/H] relation.

First we analyze the results with the enriched infall coming from dSph
galaxies.  We see that for oxygen we obtain a better agreement with
the data in the halo phase when we consider the enriched infall
models.  We recall that a key ingredient of the 2IM model is the
presence of a threshold in the gas density in the star formation (SF)
fixed at 4 M$_{\odot}$ pc$^{-2}$ in the Halo-thick disc phase.  During
the halo phase such a critical threshold is reached only at
$t=0.356\,\mathrm{Gyr}$ from the Galaxy formation.  On the other hand,
when including the environmental effect, we have to consider also the
time for the onset of the galactic wind, which in the dSph model
occurs at $t_\mathrm{gw}=0.013\,\mathrm{Gyr}$.

Therefore, the SF begins after 0.356 Gyr from the beginning of Galaxy
formation, and this fact explains the behavior of the curves with
enriched infalls in the left panel of  Fig.  \ref{O1}: during the first 0.356 Gyr in both
``2IM+dSph'' and ``2IM+dSph MIX'' models, no stars are created, and
the chemical evolution is traced by the gas accretion.
 In Figs. 2 and 3   we indicate with  thinner lines the ISM chemical evolution phases in which the SFR did not start yet in the Galactic halo, and during which stars are no created. 
 To summarize, for the  model ``2IM+dSph'' we  distinguish three different phases in the halo chemical evolution:
\begin{itemize}

\item Phase 1): 0-0.013 Gyr, the infall is primordial, the wind in dSphs has not started yet,  and there is no SF;
\item Phase 2): 0.013-0.356 Gyr, the infall is enriched by dSphs, the SFR is zero in this phase;

\item Phase 3): 0.356-1 Gyr; the infall is enriched by dSphs, the SFR is different from zero.
\end{itemize} 

During  phase 3), the SF takes over, and increases
the [O/Fe] values because of the pollution from massive stars on short
time-scales.

 We note that the entire spread of the data cannot be explained
 assuming a time dependent enriched infall with the same abundances
 of the outflowing gas from dSph galaxies, even if there is a
 better agreement with the halo data in comparison to the model with
 primordial infall.

It is important to underline that, until the SF is  non-zero, no stars are
created; however, since our models follow the chemical abundances in
the gas phase, the solely contribution to the ISM chemical evolution
before SF begins is due to the time dependent enriched infall. It means that
in the ``2IM+dSph'' model the first stars that are formed have [Fe/H]
values larger than -2.4 dex.

In this case, to explain data for stars with  [Fe/H] smaller than -2.4 dex
we need stars formed in dSph systems (see the model curve of the
chemical evolution of dSph galaxies).

Concerning the results with the enriched infall from UfD outflow
abundances, we recall here that in our reference model for UfD
galaxies, the wind starts at 0.08 Gyr. The model results for the halo
still reproduce the data but with the same above mentioned caveat.
The models with enriched infalls which show the fastest chemical
enrichment are the ones with infall abundances taken from the outflows
of dSph objects, because the galactic winds occur earlier than in UfD
systems.

In Spitoni et al. (2016) we also show the results for Mg and Si. 
 As concluded for the [O/Fe] versus [Fe/H]
abundance diagram, our reference chemical evolution models for dSph
and UfD galaxies cannot explain the observed Galactic halo data over
the entire range of [Fe/H] abundances. This rules out the hypothesis
that all Galactic halo stars were stripped or accreted in the past
from dSphs or UfDs.

In the right panel of Fig. 2, we show the results for the [Ba/Fe] versus [Fe/H]
abundance diagram.  The observational data are from  Frebel et
al. (2010), as selected and binned by Cescutti et al. (2013).  By
looking at the figure,  the 2IM model  does not provide a good
agreement with the observed data-set for
$\mathrm{[Fe/H]}<-2.5\,\mathrm{dex}$.  The initial increasing trend of
the [Ba/Fe] ratios in the 2IM model is due to the contribution of the
first Ba-producers, which are massive stars with mass in the range
$12$-$30\,\mathrm{M}_{\odot}$.

One can also appreciate that our
chemical evolution models for dSphs and UfDs fail in reproducing the
observed data, since they predict the [Ba/Fe] ratios to increase at
much lower [Fe/H] abundances than the observed data.  That is due to
the very low SFEs assumed for dSphs and UfDs, which cause the first
Ba-polluters to enrich the ISM at extremely low [Fe/H] abundances. The
subsequent decrease of the [Ba/Fe] ratios is due to the large iron
content deposited by Type Ia SNe in the ISM, which happens at still
very low [Fe/H] abundances in dSphs and UfDs. Hence, in the range
$-3.5\le \mathrm{[Fe/H]} \le -2.5\,\mathrm{dex}$, while Galactic halo
stars exhibit an increasing trend of the [Ba/Fe] versus [Fe/H]
abundance ratio pattern, UfD stars show a decreasing trend (see also
Koch et al. 2013).

In the right panel of Fig. 2, all our models involving an enriched infall from
dSphs and UfDs deviate substantially from the observed trend of the
[Ba/Fe] versus [Fe/H] abundance pattern in Galactic halo stars. Such a
discrepancy enlarges for $\mathrm{[Fe/H]}<-2.4$ dex, where those
models predict always larger [Ba/Fe] ratios than the 2IM model.
 
\subsection{The Results: the Galactic halo in the model 2IMW}

In this subsection we show the results when the time dependent
enriched infall is applied to the reference model 2IMW. In
the left panel of Fig. 3 we show the results in terms of [O/Fe] versus [Fe/H] in the
solar neighborhood. 
\begin{figure}[h]
	 \centering \includegraphics[scale=0.35]{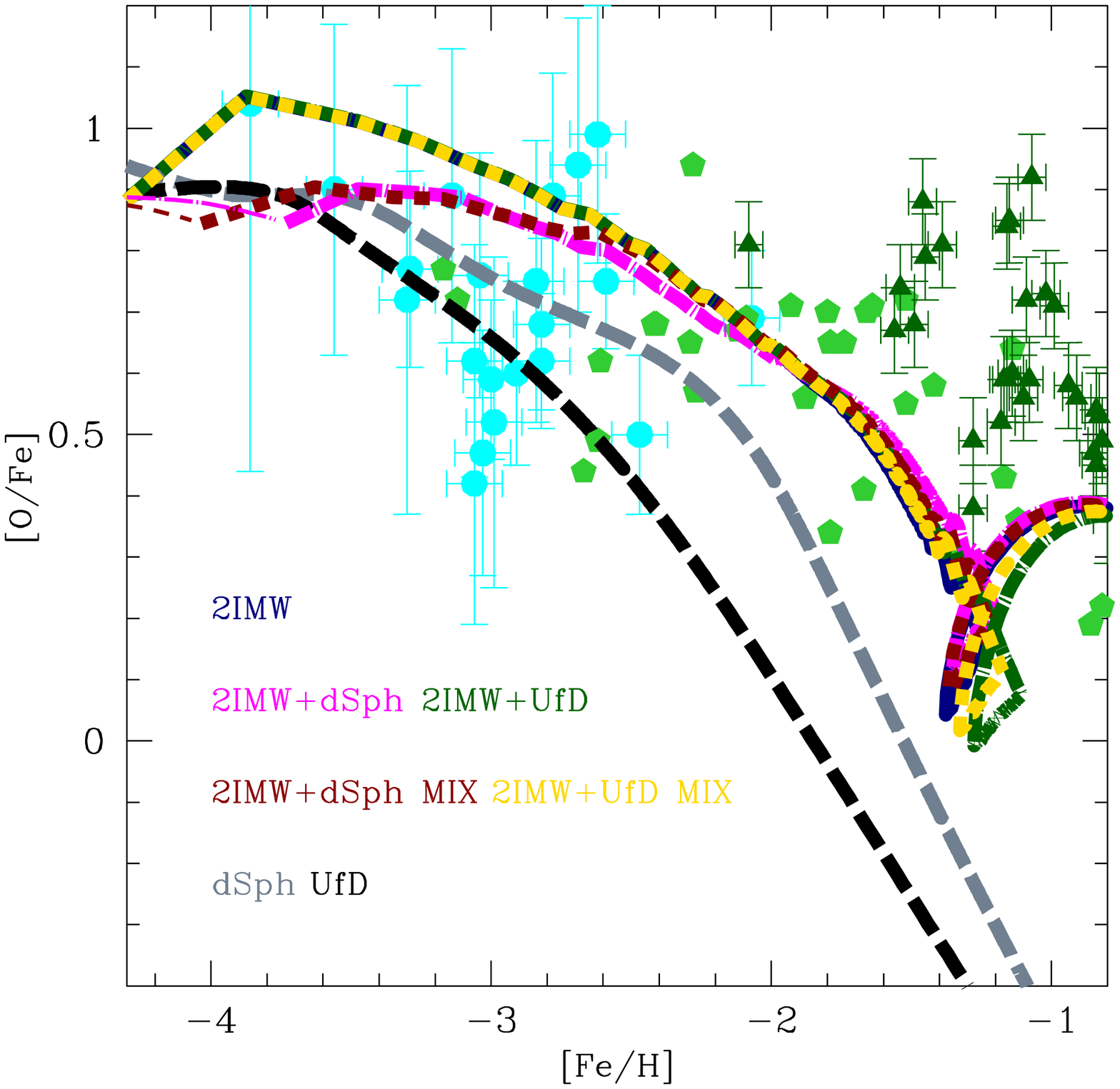}
    \centering \includegraphics[scale=0.35]{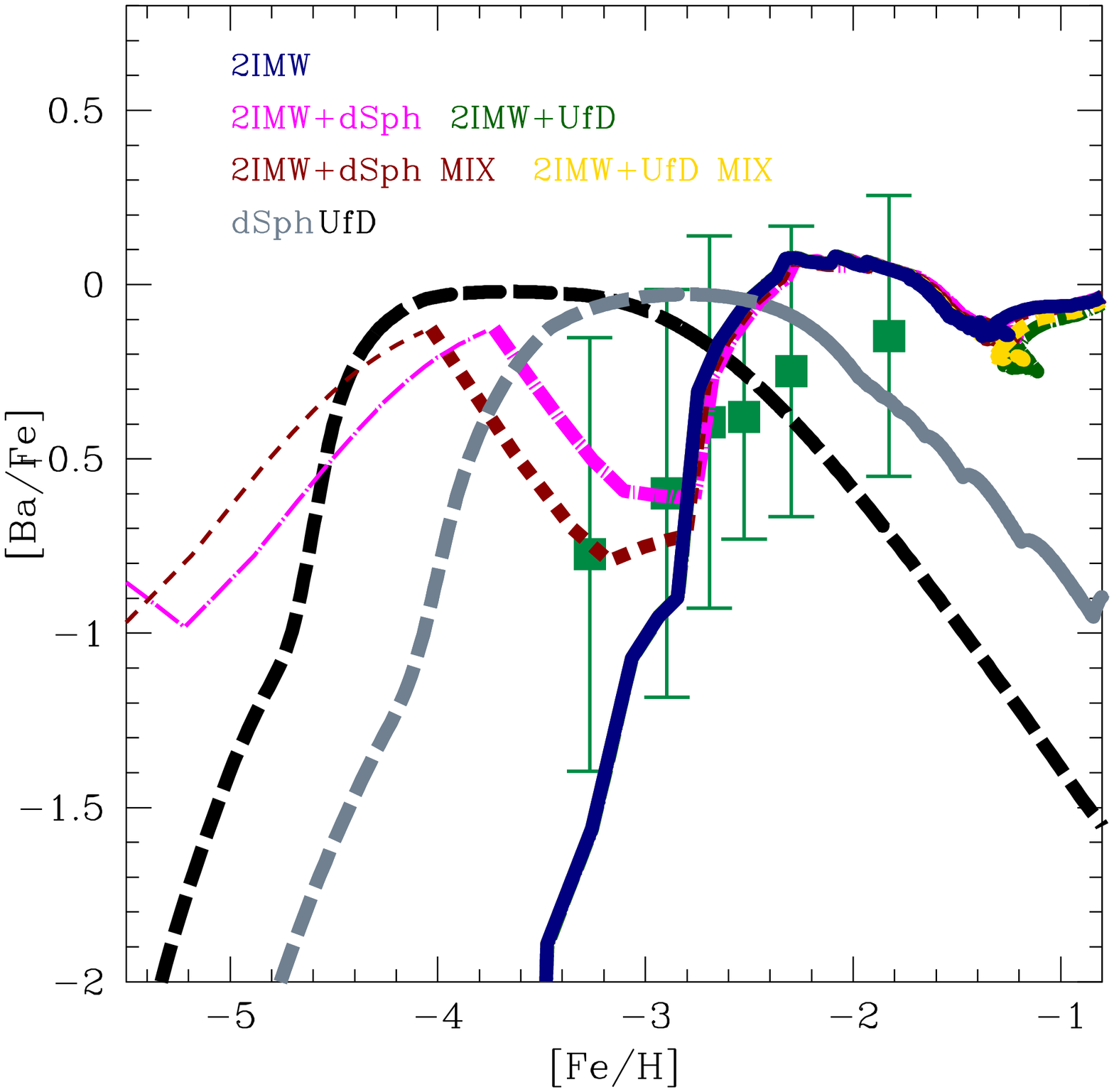} 
\caption{ As in Fig. 2  but for the 2IMW model.  
} 
\label{Ba1}
\end{figure}
In the reference model 2IMW  the SFR starts at 0.05 Gyr. Comparing  model ``2IMW+dSph'' in the left panel of Fig. 3 with  model ``2IM+dSph''
in the left panel of Fig. 2, we can see that the former  shows a   shorter
phase 2) than the latter.

The model results for the model ``2IMW+UfD'' in the left panel of Fig. 3 overlap to
the reference model 2IMW at almost all [Fe/H] abundances.  In fact,
since in the UfD galactic model the wind starts at 0.088 Gyr and, at
this instant, in the model 2IMW the SF is already active  Therefore
the effect of the enriched infall is almost negligible compared to the
pollution of chemical elements produced by dying halo stars.

Concerning the [Ba/Fe] versus [Fe/H] abundance pattern, in the right panel of Fig. 3
we compare the predictions of our models with the Galactic halo data.
We notice that the 2IMW model provides now a better agreement with the
observed data than the 2IM model, although the predicted [Ba/Fe]
ratios at $\mathrm{[Fe/H]}<-3\,\mathrm{dex}$ still lie below the
observed data.  On the other hand, by assuming an enriched infall from
dSph or UfD galaxies, the predicted [Ba/Fe] ratios agree with the
observed data also at $\mathrm{[Fe/H]}<-3\,\mathrm{dex}$.  In
conclusion, in order to reproduce the observed [Ba/Fe] ratios over the
entire range of [Fe/H] abundances, a time-dependent enriched infall in
the Galactic halo phase is required.  We are aware that  for Ba  more detailed data are needed, therefore at this stage we cannot draw firm
conclusions.

\section{Conclusions}

The main  conclusions of Spitoni et al. (2016) can be summarized as follows:

\begin{enumerate}

\item the predicted   [$\alpha$/Fe] versus [Fe/H] abundance patterns of UfD and dSph chemical evolution models deviate substantially from   the observed data of the Galactic halo stars only for [Fe/H] values larger than -2 dex;  this means that at those metallicities  the chemical evolution of the Galactic halo was different than in the satellite galaxies. On the other hand,  we notice that for Ba the
chemical evolution models of dSphs and UfDs fail to reproduce the
observational observed data of the Galactic halo stars over the whole
range of [Fe/H].

\item Concerning the chemical evolution models for the MW in the presence of  enriched gas infall we obtain that: the effects  on the [$\alpha$/Fe] versus [Fe/H] plots depend on the infall time scale for the formation of the halo and the presence of a gas threshold in the star formation. In fact, the most evident effects are present for the model 2IM, characterized by  the longest time scale of formation (0.8 Gyr), and  the  longest period without star formation activity  among  all models  presented here.

\item In general,  the enriched infall by itself is not capable to explain the  observational spread in the halo data at low  [Fe/H], in the  [$\alpha$/Fe] versus [Fe/H] plots. Moreover, in the presence of an enriched infall  we need  stars produced in situ in dSph or UfD objects  and accreted later to the Galactic halo, to explain the data at lowest [Fe/H] values.

\item The optimal element to test different theories of halo formation is barium which  is (relatively) easily measured in low-metallicity stars. In fact, we  have shown  that  the predicted  [Ba/Fe] versus [Fe/H] relation  in dSphs and UfDs is quite different than in the Galactic halo.
Moreover, the [Ba/Fe] ratio can be  
substantially influenced by the assumption of an enriched infall.  In
particular, the two infall plus outflow model can better reproduce the
data in the whole range of [Fe/H] abundances, and this is especially
true if a time dependent enriched infall during the halo phase is
assumed.

\end{enumerate}

\end{document}